\documentclass[twocolumn,showpacs,pra,aps, superscriptaddress]{revtex4}
\usepackage{amssymb}
\usepackage{amsmath}
\usepackage{graphicx}

\begin{document}

\title{Enhanced current noise correlations in a Coulomb-Majorana device}

\author{Hai-Feng L\"{u}}

\affiliation{Department of Physics, South University of Science and
Technology of China, Shenzhen 518055, China}

\affiliation{Department of Applied Physics, University of Electronic
Science and Technology of China, Chengdu 610054, China}

\author{Hai-Zhou Lu}
\email{luhz@sustc.edu.cn}

\affiliation{Department of Physics, South University of Science and
Technology of China, Shenzhen 518055, China}

\author{Shun-Qing Shen}

\affiliation{Department of Physics, The University of Hong Kong,
Pokfulam Road, Hong Kong, China}

\begin{abstract}
Majorana bound states (MBSs) nested in a topological nanowire are
predicted to manifest nonlocal correlations in the presence of a
finite energy splitting between the MBSs. However, the signal of the
nonlocal correlations has not yet been detected in experiments. A
possible reason is that the energy splitting is too weak and
seriously affected by many system parameters. Here we investigate
the charging energy induced nonlocal correlations in a hybrid device
of MBSs and quantum dots. The nanowire that hosts the MBSs is
assumed in proximity to a mesoscopic superconducting island with a
finite charging energy. Each end of the nanowire is coupled to one
lead via a quantum dot with resonant levels. With a floating
superconducting island, the devices show a negative differential
conductance and giant super-Poissonian shot noise, due to the
interplay between the nonlocality of the MBSs and dynamical Coulomb
blockade effect. When the island is strongly coupled to a bulk
superconductor, the current cross correlations at small lead
chemical potentials are negative by tuning the dot energy levels. In
contrast, the cross correlation is always positive in a non-Majorana
setup. This difference may provide a signature for the existence of
the MBSs.
\end{abstract}


\date{\today}

\maketitle

\section{Introduction}

Quantum transport through topological insulators and superconductors
has received a large amount of attention in condensed matter physics
over the past few years \cite{Hasan10RMP,Qi11RMP,ShenBook}. One of
the most influential discoveries is that topological phases
supporting Majorana fermions can be realized and engineered in the
heterostructures based on s-wave superconductors and materials with
strong spin-orbit interaction
\cite{Sau10PRL,Oreg10PRL,Lutchyn10PRL,Alicea10PRB,Sau10PRB,Beenakker13ARCMP}.
The search for Majorana bound states (MBSs) is motivated in part by
their non-Abelian characteristics and potential application in
fault-tolerant quantum computations
\cite{Nayak08RMP,Stern10Nat,Ivanov01PRL,Kitaev03AP,Moore91NPB,Fu10PRL,Flensberg11PRL,Leijnse11PRL,Alicea11NP,Jiang11PRL,Bonderson11PRL}.
It was predicted that the signature of the MBSs may exhibit as a
zero bias conductance peak in the normal metal/topological
superconductor junctions, as has been observed in hybrid devices of
superconductor and semiconductor nanowire
\cite{Mourik12Science,Deng12NL,Das12NP,Rokhinson12NP,Finck13PRL} and
in ferromagnetic iron atomic chains on the surface of
superconducting lead \cite{NadjPerge14Science}. Nevertheless, an
unambiguous experimental verification remains elusive because
zero-bias peaks can also have non-Majorana origins, such as the
Kondo effect or disorder effect
\cite{Bagrets12PRL,Pikulin12NJP,Liu12PRL,Kells12PRB,Rainis13PRB}.

The signal of current noise cross correlation could provide an
alternative, even confirmative proof, to verify the existence of
MBSs. In non-Majorana devices, it has been demonstrated
experimentally that Cooper pairs can split into spin-entangled
electrons flowing in two spatially separated normal metals,
resulting in a positive current-current correlation
\cite{Hofstetter09Nature,Hofstetter11PRL,Wei12NatPhys,Das12NatCommun}.
Although many theoretical studies have been devoted to investigate
the property of the noise cross correlation induced by MBSs
\cite{Bolech07PRL,Nilsson08PRL,Tewari08PRL,Law09PRL,Lu12PRB,Zocher13PRL,Liu13PRB,Wang13PRB,Wang13EPL,Lu14PRB},
the Majorana-modulated nonlocal transport signal has not been
reported experimentally up to now. One of the possible reasons is
that the energy splitting of MBSs, which is essential to induce the
noise cross correlation, is usually weak. The Majorana energy
splitting is at most tens of $\mu$eV in a topological nanowire with
a length of $1-2\mu$m \cite{Sarma12PRB}, and it approaches zero near
the critical Zeeman field between the trivial and topological phases
\cite{Alicea10PRB,Sarma12PRB}. Compared to the Majorana energy
splitting, other energy scales, such as the intra- and inter-dot
Coulomb interaction and superconducting pairing energy, are much
stronger. For instance, a typical value of the inter-dot Coulomb
interaction is of the order of $0.1$meV
\cite{McClure07PRL,Zhang07PRL}, and the pairing energy of a Cooper
pair is of the order of $0.1$eV. Therefore, it is highly desirable
to produce robust current-current correlations by combining the
nonlocality of Majorana fermions and other robust physical
mechanisms.

Recently, quantum transport properties modulated by Coulomb
interactions in Majorana devices have attracted much attention
\cite{Gangadharaiah11PRL,
Stoudenmire11PRB,Hutzen12PRL,Altland13PRL,Beri12PRL, Galpin12PRB}.
In an interacting transistor coupled to MBSs, it is found that the
conductance shows Coulomb oscillations with universal halving of the
finite temperature peak under strong blockade conditions
\cite{Hutzen12PRL}. Altland and Egger studied multiple helical
nanowires in proximity to a common mesoscopic superconducting island
with a finite charging energy \cite{Altland13PRL}. The MBSs prepared
in a superconducting island with Coulomb interactions are suggested
to realize the topological Kondo effect \cite{Beri12PRL,
Galpin12PRB}. B\'{e}ri and Cooper studied the simplest case with
three leads coupled to two pairs of Majorana fermions with a
charging energy \cite{Beri12PRL}. The superconducting Coulomb island
supporting MBSs offers a new playground to generate nonlocal
current-current correlations in the absence of a finite Majorana
energy splitting. However, since the large global charging energy is
itself a strong nonlocal perturbation, it becomes troublesome
whether the measured nonlocal current correlations are essentially
generated by the charging energy or MBSs. It has been pointed out
that Poissonian shot noises are generically obtained by MBSs in a
floating topological superconductor in the absence of charging
energy \cite{Ulrich15PRB}. However, it does not provide a unique
signature to confirm the existence of MBSs. Although the
Coulomb-modulated conductance properties in Majorana devices have
been well studied, it remains unknown how Coulomb interaction
affects the nonlocal current noise cross correlations in Majorana
systems.

\begin{figure}[htbp]
\centering
\includegraphics[width=0.45\textwidth]{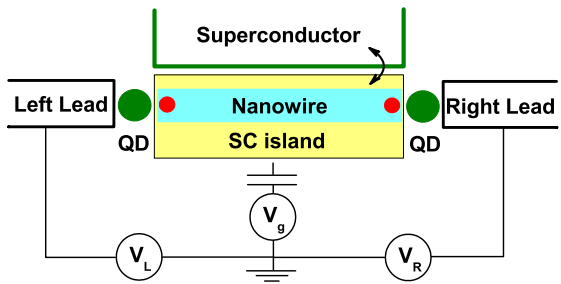}
\caption{(Color online) Schematic view
of the device. A Majorana nanowire is proximity-coupled
to a mesoscopic superconducting (SC) island with a finite charging
energy. The nanowire is in the topological superconducting phase and a pair
of MBSs (marked as the red dots) appear in the wire ends. Each end of the nanowire
is connected to a normal metal electrode via a quantum dot (QD). In
addition, the island is coupled to
another grounded bulk superconductor. The dimensionless gate parameter $n_g$ (see main text) is proportional to
a gate voltage $V_g$ that controls the average charge on the SC island.
The leads are biased with the chemical potential $V_L$ and $V_R$. Crossed Andreev
reflections can be induced by correlating the currents
that flow into the topological nanowire via the MBSs. The nonlocal current cross correlations
could be generated by Coulomb interactions in the absence of the Majorana energy splitting.}
\label{fig-1}
\end{figure}

In this paper, we investigate the nonlocal transport modulated by
the Coulomb interactions in devices comprising a Majorana nanowire
contacted to quantum dots and leads, where the nanowire is in
proximity to a mesoscopic superconducting island with a finite
charging energy. The purpose to insert the quantum dots is to offer
an efficient way to modulate the nonlocal current correlations
without affecting the property of the MBSs. The paper is organized
as follows. In Sec. II, we introduce the model Hamiltonian of the
Majorana-dot device fabricated on a superconducting Coulomb island,
as well as the current and noise cross correlation formulas. In Sec.
III, we investigate the nonlocal transport properties modulated by
the charging energy in the island. We consider the cases that the
superconducting island is floating and connected to a bulk
superconductor, respectively. For comparison, the current noise
cross correlation property in the absence of MBSs, is also
discussed. Finally, a summary is given in Sec. IV.

\section{Model and formalism}

\subsection{Model Hamiltonian}
A schematic of the Coulomb-Majorana junction is illustrated in Fig.
1. We consider a semiconductor nanowire proximity-coupled to a
mesoscopic superconducting island with a finite charging energy, and
there exists a pair of MBSs at the wire ends under a proper magnetic
field \cite{Beenakker13ARCMP}. Each end of the nanowire connects to
a normal metal electrode via a quantum dot. The intraisland Coulomb
interactions introduce correlations between the two MBSs, and
thereby generate the cross correlation between the currents flowing
through the two quantum dots. It is convenient to define nonlocal
auxiliary fermion operators $\eta_{a}=f+f^{\dagger}$,
$\eta_{b}=i(f^{\dagger}-f)$ for the MBSs, with the number operator
$\hat{n}_f=f^\dagger f$. In the regular fermionic representation,
the instantaneous charged state of the superconducting island is
described by ($N_c, n_f$), where the integer $N_c$ represents the
Cooper pair number in the island and $n_f$ is the eigenvalue of
$\hat{n}_f$, respectively. The total Hamiltonian is given by
\begin{eqnarray}
H &=& H_w+H_d+H_c+H_b+H_l+H_t.
\end{eqnarray}
The nanowire is in the topological superconducting state and
isolated MBSs appear at the wire ends. Including the Majorana energy
splitting and a Coulomb charging term, the island Hamiltonian is
fully expressed by \cite{Hutzen12PRL}
\begin{eqnarray}\label{H-w}
H_w =\varepsilon_{M}(f^{\dagger}f-1/2)+E_c(2 \hat{N}_c + \hat{n}_f- n_g)^2,
\end{eqnarray}
where $\varepsilon_{M}$ denotes the Majorana energy splitting and
the capacitive charging effect is denoted by $E_c$ and can be tuned
by a gate voltage parameter $n_g\propto V_g$. Experimentally, the
charging energy in a superconducting island can reach at about
$100\mu$eV \cite{Sillanpaa05PRL,Naaman07PRL}. Next, the two quantum
dots in Fig. 1 are assumed to be in the Coulomb blockade regime such
that each dot can be modeled as a single fermion level. The energy
level in each dot can be tuned by gate voltages, and the spin
degeneracy on the dots also breaks by the Zeeman field which induces
the topological superconducting phase of the wire. As a consequence,
we can use the effectively spinless fermion operators
$d_{j}^{\dagger}$ ($d_{j}$) for the dots
\cite{Zazunov11PRB,Zazunov12PRB,Plugge15PRB}. The Hamiltonian of the
quantum dots reads
\begin{eqnarray}
H_{d}=
\varepsilon_{1}d_{1}^{\dagger}d_{1}+\varepsilon_{2}d_{2}^{\dagger}d_{2},
\end{eqnarray}
where $\varepsilon_{j}$ denotes the dot energy level and
$d_{j}^{\dagger}$ ($d_{j}$) represents the electron creation
(annihilation) operator on dot $j$.

Different from the noninteracting case in which $E_c=0$, there
exists an energy cost when absorbing or emitting a Cooper pair in
the superconducting island. To ensure the charge conservation, it is
essential to add a factor $e^{\mp i\varphi}$ in the tunneling terms
$d_j^\dagger f^\dagger$ ($d_jf$), which changes the Cooper pair
number $N_c$ by one unit and thus restores the charge balance. The
Hamiltonian of the Majorana-dot tunneling then reads
\cite{Zazunov11PRB,Zazunov12PRB,Plugge15PRB}
\begin{eqnarray}\
H_{c} =\lambda_{1}d_1^{\dagger}(e^{-i\varphi}f^{\dagger} +f)
+ \lambda_{2}(e^{-i\varphi}f^{\dagger} -f )d_2^{\dagger}+h.c.,
\end{eqnarray}
where $\lambda_{1,2}$ denotes the dot-Majorana coupling strength and
the operator $e^{\pm i\varphi}$ raises (lowers) the Cooper pair
number by one unit, i.e., $N_c\rightarrow N_c\pm1$. In conjunction
with spin-orbit coupling and the Zeeman splitting, the nanowire in
the topological phase resembles an effective $p$-wave
superconductor. Therefore, the Cooper pair exchange with the bulk
superconductor involves the spin-flip scattering processes. The spin
properties of the topological nanowire as well as spin-flip
processes in the contact can be fully taken into account via the
tunnel couplings $\lambda_{1,2}$ and $E_J$, where $E_J$ denotes the
Cooper pair exchange strength between the nanowire and the bulk
superconductor. The tunnel couplings capture the possible spin
dependence of microscopic transition amplitudes and can be taken as
real-valued positive \cite{Zazunov11PRB}. We include the Cooper pair
exchange between the superconducting island and another bulk
superconductor,
\begin{eqnarray}
H_b=-E_J\cos \varphi,
\end{eqnarray}
with the Josephson coupling $E_J$.

Finally, the electrodes and the dot-electrode
tunneling are described by the Hamiltonians
\begin{eqnarray}
H_l&=&\sum_{j k }\varepsilon_{jk} a_{jk }^\dagger a_{jk},
\nonumber\\
H_t&=&\sum_{j k }\left(t_j a_{jk }^\dagger d_{j}+h.c.\right),
\end{eqnarray}
where $a_{jk }^\dagger$ ($a_{jk }$) is the electron creation
(annihilation) operator in the lead $j$ with an energy
$\varepsilon_{jk}$ and $t_j$ is the lead-dot coupling strength.

The current noise cross correlation has been well studied in a
dot-Majorana-dot structure when the Majorana nanowire is directly in
proximity with a bulk superconductor
\cite{Lu12PRB,Zocher13PRL,Liu13PRB,Wang13PRB,Wang13EPL,Lu14PRB}. In
this case, a finite Majorana energy splitting is essential to
generate nonlocal correlations. In contrast, here the cross
correlations could be controlled by the charging energy of the
superconducting island as well as the Majorana energy splitting. If
we consider a one-dimensional tight-binding Kitaev model
\cite{Kitaev03AP}, the two quantum dots look like two extended sites
of the Kitaev model. In this sense, the Majorana zero modes will be
located near the interface between the nanowire and the quantum
dots, where the tail part of the wave function of the zero modes
could enter the quantum dots. Differently, the dot energy levels
could be tunable by applying gate voltages. If the dot energy levels
are tuned to be high enough, the wave function of Majorana mode will
appear hardly at the quantum dots.

\subsection{Diagonalized master equation approach}
We exploit the diagonalized master equation (DME) approach to
investigate the electronic transport through this system in the
sequential tunneling regime
\cite{Bruus04Book,Kashcheyevs06PRB,Mitra07PRB,Timm08PRB,Poltl09PRB,Lu09PRB}.
In our previous paper \cite{Lu14PRB}, the applicability of the DME
approach has been discussed in the Majorana devices, by comparing
with the nonequilibrium Green's function (NEGF) method. It is shown
that the DME works well in most regimes of system parameters, while
it breaks down for strong central region-lead coupling or when
energy degeneracies appear. Here we extend our discussion to the
interacting case. Compared to the NEGF method, the DME approach is
convenient in dealing with many interacting energy levels, and there
is strong coherence between different levels.

In the DME approach, we firstly diagonalize the Hamiltonian of the
island-dot part and obtain the eigenvalues $E_n$ and their
corresponding eigenfunctions $|\beta_n\rangle$. Different from the
case that the superconductor is grounded directly, a superconducting
island with a finite charging energy is used in the proposed
tunneling device. To ensure the charge conservation, the total
Cooper pair number $N_c$ should also be considered as a degree of
freedom in the calculation \cite{Plugge15PRB}. To give an explicit
matrix form of the island-dot Hamiltonian, it is convenient to index
the states with the occupation numbers
\begin{eqnarray}
|n_1n_2n_{f}N_c\rangle=(d_1^{\dagger})^{n_1}(d_2^{\dagger})^{n_2}
(f^{\dagger})^{n_{f}}|000N_c\rangle.
\end{eqnarray}
Here the quantum numbers
can take the values $n_j, n_{f}=0,1$ and $N_c=-N_{m},...N_{m}$, where $N_m$
is the cutoff for the Cooper pair number.

In the Born-Markov approximation, the time evolution of the density
matrix $\rho_{D}(t)=\{|\beta_{n}\rangle\langle\beta_{n'}|\}$ in
terms of the states $|\beta_n\rangle$ is given by the rate equations
\begin{eqnarray}\label{RateEq}
\frac{d}{dt}\rho_{D}(t)=\mathbf{W}\rho_{D}(t),
\end{eqnarray}
where the elements of the rate matrix are given
by \cite{Lu12PRB,Poltl09PRB}
\begin{eqnarray}
W_{n'n} & = & \sum_{j}\Gamma_{j}\left[f(\Delta_{n'n}+\mu_{j})|\langle\beta_{n'}|d_{i}|\beta_{n}\rangle|^{2}\right.\nonumber \\
 &  & \left.+f(\Delta_{n'n}-\mu_{j})|\langle\beta_{n'}|d_{j}^{\dagger}|\beta_{n}\rangle|^{2}\right]
\end{eqnarray}
for $n\neq n'$, and
\begin{eqnarray}
W_{nn} & = & -\sum_{n'\neq n}^{N}W_{n'n}.
\end{eqnarray}
Here $f_{j}(\omega)=[1+e^{\omega/k_{B}T}]^{-1}$ is the Fermi-Dirac
distribution function, $\mu_{j}$ is the chemical potential in lead
$j$, and $\Delta_{k'k}$ is the Bohr frequency of the transition from
$|\beta_{k}\rangle$ to $|\beta_{k'}\rangle$. In the wide-band limit
approximation, the dot-lead coupling for dot $j$ is measured by the
parameter $\Gamma_{j}=2\pi|t_{j}|^{2}\rho_{j}$, with $\rho_{j}$ the
spinless density of states near the Fermi surface of lead $j$.

The steady-state current $I_{j}$ is given by
\begin{eqnarray}
I_{j} & = & \sum_{k}[\hat{\mathbf{\Gamma}}^{j}\rho_{D}^{(0)}]_{k},
\end{eqnarray}
where $\rho_{D}^{(0)}$ is the steady-state solution of Eq.
(\ref{RateEq}), $\hat{\mathbf{\Gamma}}^{j}$ is the matrix of the
current operator and its elements are given by
\begin{eqnarray}
\hat{\Gamma}_{k'k}^{j} & = & \Gamma_{j}\left[f(\Delta_{k'k}+\mu_{j})|\langle\beta_{k'}|d_{j}|\beta_{k}\rangle|^{2}\right.\nonumber \\
 &  & \left.-f(\Delta_{k'k}-\mu_{j})|\langle\beta_{k'}|d_{j}^{\dagger}|\beta_{k}\rangle|^{2}\right].
\end{eqnarray}
The first and the second terms of $\hat{\Gamma}_{k'k}^{j}$ represent
the tunneling amplitudes flowing into and out of the lead,
respectively.

We are focusing on the current noise correlations modulated by the
charging energy on the superconducting island. It is well known that
the noise power spectra can be expressed as the Fourier transform of
the current-current correlation function
\begin{eqnarray}
S_{I_{i}I_{j}}(\omega) & = & 2\langle I_{i}(t)I_{j}(0)\rangle_{\omega}-2\langle I_{i}\rangle_{\omega}\langle I_{j}\rangle_{\omega}.
\end{eqnarray}
Here, $I_{i}$ and $I_{j}$ are the electrical currents across dot $i$
and dot $j$, respectively, and $t$ is the time. Furthermore, the
current-current correlation function of the currents $I_{i}$ and
$I_{j}$ can be expressed in terms of the density matrix as
\begin{eqnarray}
\langle{I_{i}(t)I_{j}(0)}\rangle & = & \theta(t)\sum_{k}[\hat{\Gamma}^{i}\hat{T}(t)\hat{\Gamma}^{j}{\bm{\rho}}^{(0)}]_{k}\nonumber \\
 &  & +\theta(-t)\sum_{k}[\hat{\Gamma}^{j}\hat{T}(-t)\hat{\Gamma}^{j}{\bm{\rho}}^{(0)}]_{k},
\end{eqnarray}
with $\hat{T}(t)=\exp[\mathbf{W}t]$ the propagator governing the
time evolution of the density matrix element $\rho_{k}(t)$. Finally,
the current-current correlation in the $\omega$-space becomes
\begin{eqnarray}
\langle I_{i}(t)I_{j}(0)\rangle_{\omega} & = & \sum_{k}
\left[\hat{\Gamma}^{i}\hat{T}(\omega)\hat{\Gamma}^{j}
{\bm{\rho}}^{(0)}+\hat{\Gamma}^{j}\hat{T}(-\omega)
\hat{\Gamma}^{i}{\bm{\rho}}^{(0)}\right]_{k},\nonumber \\
\end{eqnarray}
where $\hat{T}(\pm\omega)=\left(\mp
i\omega\hat{I}-\mathbf{W}\right)^{-1}$.

\section{Numerical results and discussion}

In the following, we present numerical results for the sequential
tunneling currents and their correlations in the presence of the
charging energy on the superconducting island in the nonlinear
response regime. The negative noise cross correlation is related to
the antibunching between tunneling events, which arises as a result
of the Pauli exclusive principle of scattered electrons, while the
positive cross correlation is usually related to the bunching of
tunneling processes, e.g., due to the interchannel Coulomb blockade
or crossed Andreev reflection
\cite{Hofstetter09Nature,Hofstetter11PRL,Wei12NatPhys,Das12NatCommun}.
In the calculation, we adopt the symmetric coupling strength as
$\Gamma_{1,2}=\Gamma_0$ and $\lambda_{1,2}=10\Gamma_0$, where
$\Gamma_0$ serves as a convenient energy unit. The energy levels
$\varepsilon_{1,2}$ of the quantum dots are tunable by applying gate
voltages.

For comparison, we separately discuss two different cases: one is
with $E_J=0$ and the other is with $E_J\gg\Gamma_0, \lambda_0,
k_BT$. The case of $E_J=0$ corresponds to a floating superconducting
island, and the electrons in the island can only tunnel through the
quantum dots. For $E_J=0$, the current flows from the left lead to
the right lead when a bias voltage $V_b$ is applied and we take
$\mu_L=-\mu_R=V_b/2$ in the calculation. For $E_J\gg\Gamma_0,
\lambda_0, k_BT$, the island is strongly coupled to the bulk
superconductor. In this case, we take $\mu_L=\mu_R=V_0$ and the
currents flow from the two electrodes into the superconducting
island. The sign of the current noise cross correlation sensitively
depends on the interplay between the intraisland Coulomb interaction
and the crossed Andreev reflection. In the limit that
$E_C\rightarrow0$ and $E_J\rightarrow\infty$, the result should
reduce to the case when the bulk superconductor is grounded, which
has been well studied previously
\cite{Lu12PRB,Zocher13PRL,Liu13PRB,Wang13PRB,Wang13EPL,Lu14PRB}.

\subsection{Floating superconducting island ($E_J=0$)}

\begin{figure}[htbp]
\centering
\includegraphics[width=0.45\textwidth]{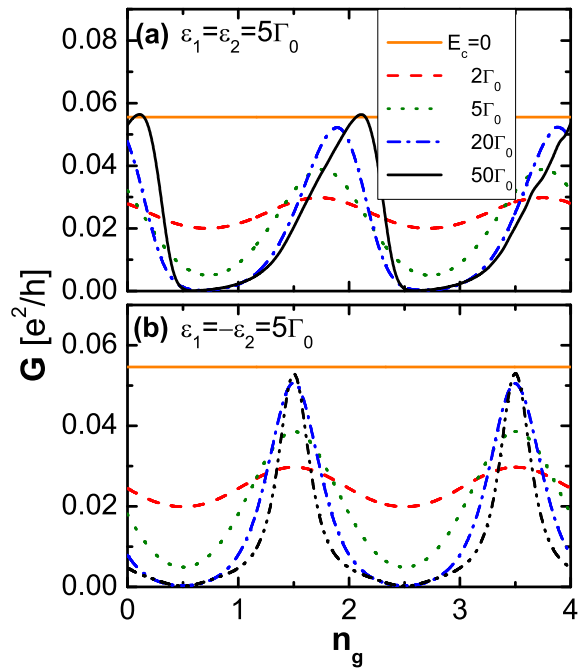}
\caption{The linear conductance $G$
as a function of $n_g$
for different charging energies $E_c$ on the superconducting island. (a) The symmetric configuration of
dot energy levels ($\varepsilon_1=\varepsilon_2$). (b) The antisymmetric dot level configuration ($\varepsilon_1=-\varepsilon_2$). Other parameters:
$E_J=0$, $k_{B}T=2\Gamma_0$, $\varepsilon_{1}=5\Gamma_0$, $\varepsilon_{M}=0$, $\lambda_{1,2}=10\Gamma_0$, and
$\Gamma_0$ is taken as the energy unit.}
\label{fig-2}
\end{figure}

Firstly we address the $n_g$ dependence of the linear conductance
($V_b\rightarrow0$), as shown in Fig. \ref{fig-2}(a) for the
symmetric dot energy level configuration
($\varepsilon_1=\varepsilon_2$), and in Fig. \ref{fig-2}(b) for the
antisymmetric level configuration ($\varepsilon_1=-\varepsilon_2$).
In both configurations, the conductance shows clear $n_g$ dependent
oscillations in addition to a constant part. With the increase of
the charging energy $E_c$, the constant part is considerably
suppressed while the oscillation amplitude increases. As shown in
the charging part of $H_w$ in Eq. (\ref{H-w}), a shift
$n_g\rightarrow n_g \pm 2$ can be compensated by absorbing or
emitting a Cooper pair $N_c \rightarrow N_c \pm 1$. Therefore, all
observables are periodic in $n_g$ with a period of $\Delta n_g= 2$.
This feature is different from the case when the MBSs are directly
coupled to the two electrodes, in which a zero-energy Majorana mode
($\varepsilon_M=0$) is considered and the oscillation period is
$\Delta n_g= 1$ \cite{Hutzen12PRL}. When Majorana fermions are
directly coupled to two leads, the parity change in the
superconducting island costs a finite energy and the period becomes
$\Delta n_g= 2$ for $\varepsilon_M\neq0$. When two quantum dots are
inserted in, the period is $\Delta n_g= 2$ for either
$\varepsilon_M=0$ or $\varepsilon_M>0$. In Fig. 2, it is indicated
that the value of conductance is much smaller than $e^2/h$. This is
because a relatively high temperature ($k_BT=2\Gamma_0$) is adopted
in the calculation. In this case, the conductance would be
considerably suppressed compared to the case of zero temperature
\cite{Hutzen12PRL}.

Another feature in Fig. \ref{fig-2} is that the conductance curves
are symmetric about $n_g=1/2$ for the antisymmetric dot level
configuration, which is absent for the symmetric case
$\varepsilon_1=\varepsilon_2$. By applying a particle-hole
transformation that exchanges the creation and annihilation
operators, the total Hamiltonian is invariant under the replacement
$n_g\rightarrow 1-n_g$ for $\varepsilon_1=-\varepsilon_2$. Due to
the periodicity, the observables are symmetric about $n_g=2n+1/2$,
where $n$ is an integer.

\begin{figure}[htbp]
\centering
\includegraphics[width=0.45\textwidth]{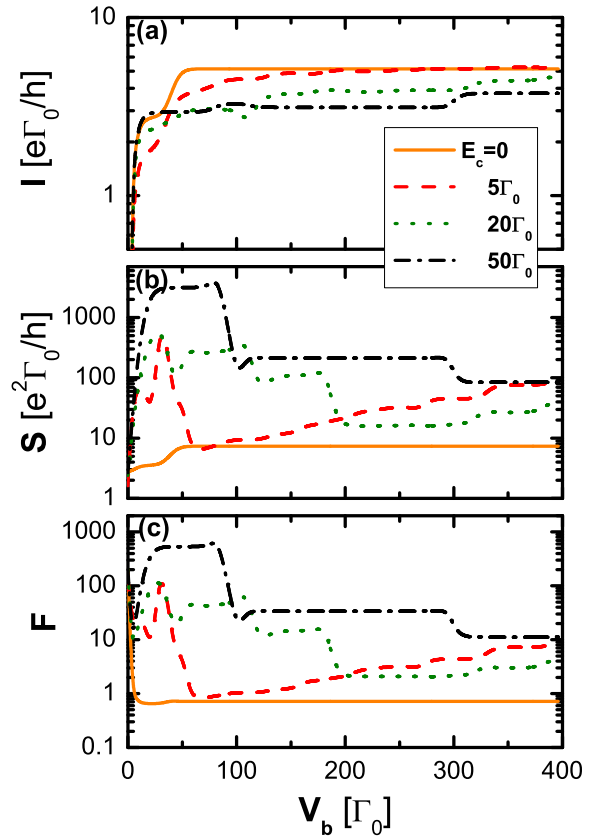}
\caption{(a) The current $I$, (b) the shot noise $S$ and (c)
the Fano factor $F$ as functions of the bias voltage
$V_b$ for different charging energies $E_c$. Other parameters:
$E_J=0$, $k_{B}T=2\Gamma_0$, $\varepsilon_{1}=\varepsilon_{2}=5\Gamma_0$,
$\varepsilon_{M}=0$, $\lambda_{1,2}=10\Gamma_0$, and $n_g=0$.}
\label{fig-3}
\end{figure}

The effects of the charging energy on the transport properties are
presented in Fig. \ref{fig-3}. Due to the current conservation
$I_1=-I_2$, the current correlations in a floating two-terminal
setup obey the relationship $S_{11} = S_{22} = -S_{21} = -S_{12} =
S$. Compared to the noninteracting case, the most important
difference in the transport is that the Coulomb interaction could
induce a negative differential conductance (NDC) and strongly
enhance the shot noise. As the bias voltage increases from $V_b=0$,
the current appears and increases gradually when the voltage sweeps
through the lowest positive eigenenergy of the central region.
However, as the lead voltage increases and other energy levels lie
in the transport window, the charging energy induces a strong
competition between different tunneling paths, leading to a decrease
of the current. In Fig. \ref{fig-3}(a), the current shows an
oscillating behavior for small charging energies. As the bias
voltage $V_b$ increases, the Cooper pair states $|N_c\rangle$ are
involved in the transport one by one, resulting in the current
oscillation. A similar behavior can also be observed in the bias
voltage dependence of the shot noise, as shown in Fig.
\ref{fig-3}(b). A giant shot noise $S$ can be generated in the
Coulomb blockade regime. One usually introduces the Fano factor
$F=S/2eI$ to represent the deviation from Poissonian shot noise for
which $F=1$. The noise Fano factor $F$ is demonstrated as a function
of the bias voltage in Fig. \ref{fig-3}(c). In the noninteracting
case $E_c=0$, the shot noise is always a sub-Poissonian type, i.e.,
$F<1$. In the presence of a finite charging energy, the
super-Poissonian shot noise ($F>1$) is induced in most transport
regimes.

\begin{figure}[htbp]
\centering
\includegraphics[width=0.48\textwidth]{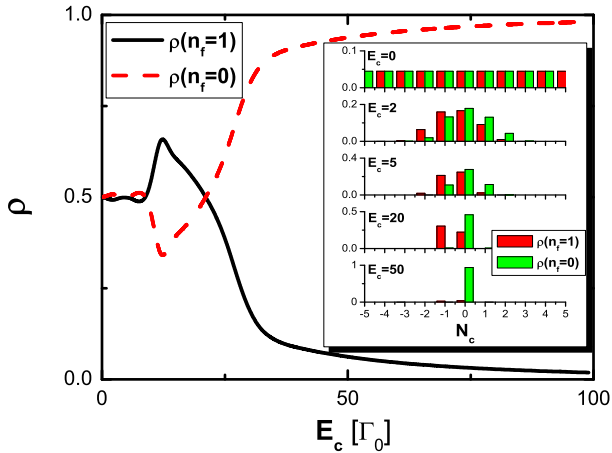}
\caption{The state population $\rho(n_f)$ in the island as a function of the charging energy.
The inset shows the the distribution of $\rho(n_f)$ as a function of the Cooper pair number $N_c$.
For $E_c=0$, all the states $|n_f, N_c\rangle$ in the island have the equal probability and we take a cutoff
of the Cooper pair number ($N_c\in [-5,5]$).
Other parameters: $V_b=30\Gamma_0$,
$E_J=0$, $k_{B}T=2\Gamma_0$, $\varepsilon_{1}=\varepsilon_{2}=5\Gamma_0$,
$\varepsilon_{M}=0$, $\lambda_{1,2}=10\Gamma_0$, and $n_g=0$.}
\label{fig-4}
\end{figure}

The NDC and the giant shot noise $S$ arise from the same mechanism,
known as the dynamical channel blockade effect. Different from the
Schottky noise, which is independent of the frequency, the shot
noise reflects the dynamical tunneling correlations and is
frequency-dependent. In the case of a floating superconducting
island ($E_J = 0$), the electrons in the island can only tunnel
through the quantum dots. In this case, the superconducting island
behaves like a normal quantum dot with multiple interacting energy
levels. The giant shot noise has been demonstrated in an interacting
quantum dot with multiple tunneling channels in several previous
studies
\cite{Cottet04PRL,Wabnig09PRL,Dubrovin07PRB,Thielmann05PRB,Aghassi06PRB,Elste06PRB,Weymann08PRB,Weymann11PRB}.
The reason is traced to a dynamical channel blockade of the
mechanically aided shuttle current that occurs in devices with
highly polarization or asymmetry of the channel-lead coupling
strengths. In Fig. \ref{fig-4} we present the dependence of state
populations in the superconducting island on the charging energy
$E_c$ to demonstrate the coupling asymmetry. By tracing out the
freedoms of $n_{1,2}$ in two quantum dots, one can obtain the state
distribution $\rho(n_f, N_c)$ in the island. For $E_c=0$, all the
states $|n_f, N_c\rangle$ in the island have equal probability and
$\rho(n_f=1)=\rho(n_f=0)=1/2$. In the presence of the charging
energy, the eigenenergies of the Majorana-dot device become $E_c$
dependent and the degeneracy is lifted. This dependence remarkably
modifies the electronic occupations in comparison with the
noninteracting case. For the strong charging energy, the island
prefers to occupy the state $| n_f=0, N_c=0\rangle$ and the
occupation probabilities of other states are strongly suppressed.
With the increase of bias voltage, more eigenenergy levels of
Majorana-dot part enter the transport window. The long time
occupation of the state $| n_f=0, N_c=0\rangle$ impedes the entry of
electrons into the island region through other channels, leading the
suppression of the current and enhancement of shot noise. This
mechanism has been illustrated in several quantum transport systems,
such as multilevel quantum dot devices
\cite{Thielmann05PRB,Aghassi06PRB,Elste06PRB,Weymann08PRB,Weymann11PRB},
Franck-Condon blockade in single molecules \cite{Koch05PRL}, and
nanoscale oscillators \cite{Hubener07PRL}.

The picture presented above is based on the sequential tunneling
regime. In the Coulomb blockade regime, the first-order tunneling is
exponentially suppressed, while higher-order processes can
contribute a small current. For a weak dot-lead coupling strength
and high temperature, the current contributed by higher-order
tunneling processes is rather small, which cannot remove the
interchannel blockade remarkably. To elucidate the role of
higher-order tunneling processes, a relatively high temperature
($k_BT=2\Gamma_0$) has been considered in our calculation, in which
the current induced by thermal fluctuations is activated. In this
case, the current contributed by higher-order tunneling processes is
overwhelmed by thermal currents.

\subsection{Strong coupling between superconducting island and a bulk superconductor ($E_J\gg \Gamma_0$)}
Now we consider a three-terminal case in which the superconducting
island is strongly coupled to the bulk superconductor. In this case,
the Cooper pairs can tunnel freely between the island and the bulk
superconductor. All states $|N_c\rangle$ are strongly mixed. To
investigate the current noise cross correlation modulated by the
charging energy, symmetric chemical potentials $\mu_1=\mu_2 =V_0$
are applied to the two electrodes, and the currents flow from the
two electrodes to the superconducting island.

\begin{figure}[htbp]
\centering
\includegraphics[width=0.45\textwidth]{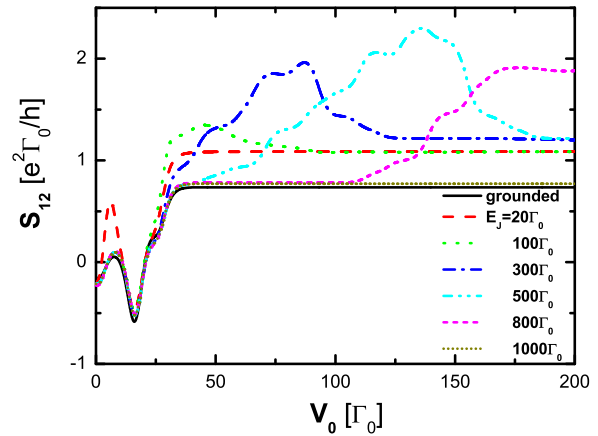}
\caption{The zero-frequency current noise
cross correlation $S_{12}$ as a function of the lead voltage
$V_0$ for different $E_J$ at $E_c=0$.
The cross correlation is induced by a finite Majorana energy splitting $\varepsilon_{M}=10\Gamma_0$.
The black line corresponds to the case in which the semiconductor nanowire is in proximity to the grounded superconductor.
The result denoted by the black line is exact and obtained by using the nonequilibrium Green's function method.
Other parameters:
$k_{B}T=2\Gamma_0$, $\varepsilon_{1}=\varepsilon_{2}=5\Gamma_0$, $\lambda_{1,2}=10\Gamma_0$, and $n_g=0$.}
\label{fig-5}
\end{figure}

In the noninteracting case where $E_c = 0$, the degree of freedom of
Cooper pair number $N_c$ decouples from those of the fermionic part
$|n_1 n_2 n_f\rangle$ in the limit $E_J\rightarrow\infty$.
Correspondingly, the transport properties reduce to the results for
a noninteracting dot-MBS-dot structure, which has been well
investigated previously
\cite{Lu12PRB,Zocher13PRL,Liu13PRB,Wang13PRB,Wang13EPL,Lu14PRB}. To
verify this physical picture, we present the current noise cross
correlation $S_{12}$ for different $E_J$, as demonstrated in Fig.
\ref{fig-5}. For comparison, we also show the results when the
semiconductor nanowire is in proximity to a grounded superconductor.
We take $E_c=0$ and $\varepsilon_M=10\Gamma_0$ to ensure that the
noise cross correlation is only induced by the Majorana energy
splitting. For the grounded case, the cross correlation $S_{12}$ is
exactly solved by using the NEGF method \cite{Lu14PRB}. According to
Fig. \ref{fig-5}, even for $E_J=50\varepsilon_M$, $S_{12}$ shows a
quite different behavior, compared to that in the directly grounded
case. However, when $E_J$ is of the order of $100\varepsilon_M$, the
cross correlations in the two cases are in good agreement with each
other in all regimes of the lead chemical potentials. Above this
strength of $E_J$, the superconducting island can be regarded as
being grounded directly.

The discussion above serves as a ground to extend the discussion to
the situation in which the current cross correlation is purely
induced by the charging energy in the superconducting island. The
Hamiltonian of the central region is exactly diagonalized in the DME
approach. The approximation of the DME approach is that the dot-lead
coupling strength is taken as a perturbation parameter and is much
weaker than other system energy scales.

\begin{figure}[htbp]
\centering
\includegraphics[width=0.45\textwidth]{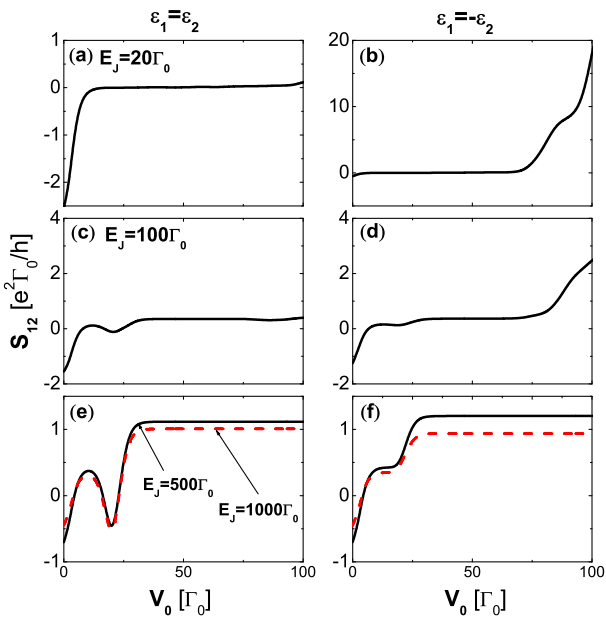}
\caption{The zero-frequency current noise
cross correlation $S_{12}$ as a function of the lead chemical potentials
$\mu_1=\mu_2=V_0$ for different $E_J$ at $\varepsilon_M=0$. We consider the
symmetric configuration
of the quantum dot levels in (a) $E_J=20\Gamma_0$, (c) $E_J=100\Gamma_0$,
(e) $E_J=500\Gamma_0, 1000\Gamma_0$ and the antisymmetric level configuration
in (b) $E_J=20\Gamma_0$, (d) $E_J=100\Gamma_0$,
(f) $E_J=500\Gamma_0, 1000\Gamma_0$, respectively.
The cross correlation is induced by the charging energy $E_c=30\Gamma_0$.
For strong couplings with the bulk superconductor, such as $E_J=500\Gamma_0$ or $1000\Gamma_0$, the Coulomb blockade
induced by the charging energy is smeared out. This case is equivalent
to the case that the superconducting island is grounded directly. Other parameters:
$k_{B}T=2\Gamma_0$, $\varepsilon_{1}=5\Gamma_0$, $\lambda_{1,2}=10\Gamma_0$, and $n_g=0$.}
\label{fig-6}
\end{figure}

In Fig. \ref{fig-6}, we demonstrate the current noise cross
correlation $S_{12}$ as a function of the lead chemical potentials
$\mu_1=\mu_2=V_0$. To ensure that the cross correlation is induced
only by the charging energy, we take $\varepsilon_M=0$ and
$E_c=30\Gamma_0$. We consider the symmetric
($\varepsilon_1=\varepsilon_2$) and antisymmetric
($\varepsilon_1=-\varepsilon_2$) configurations of the dot levels.
For the cross correlation induced by the Majorana energy splitting,
$S_{12}$ is always positive in the antisymmetric level configuration
\cite{Lu12PRB,Zocher13PRL}. However, Fig. \ref{fig-6} shows that the
current cross correlation $S_{12}$ modulated by the charging energy
is always negative in both level configurations at small lead
chemical potentials. For small $V_0$, the device lies in the Coulomb
blockade regime. The transport through the two dots compete with
each other, leading to a negative $S_{12}$. With the increase of
$V_0$, more eigenenergy levels of the dot-Majorana island-dot part
are involved in the transport and the Coulomb blockade is partly
relieved. The crossed Andreev reflection becomes dominant gradually,
resulting in a positive cross correlation. The coupling $E_J$
between the island and the bulk superconductor helps to remove the
dynamical Coulomb blockade. For a strong $E_J$, the Coulomb blockade
induced by the charging energy is smeared out. This case is
equivalent to the case in which the superconducting island is
grounded directly. For a small $E_J=20\Gamma_0$, Fig. \ref{fig-6}(b)
shows that the current cross correlation is strongly enhanced, due
to the nonlocality of the MBSs and the dynamical channel blockade
effect. Such an enhancement of the current cross correlation
provides a new signature for the existence of the MBSs.

\begin{figure}[htbp]
\centering
\includegraphics[width=0.45\textwidth]{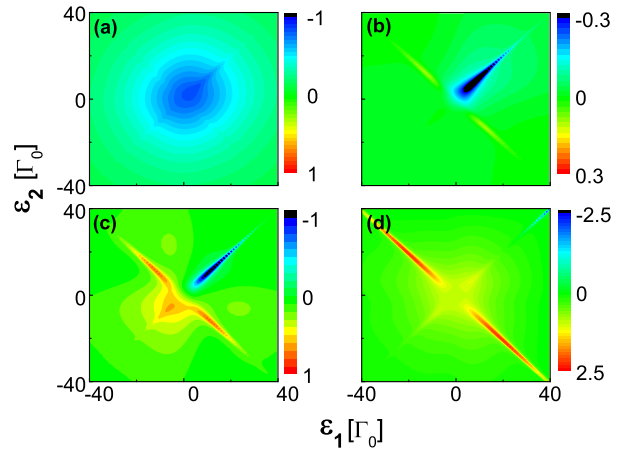}
\caption{The zero-frequency current noise
cross correlation $S_{12}$ as a function of
the dot energy levels $\varepsilon_1$ and $\varepsilon_2$
for different lead chemical potentials $\mu_1=\mu_2=V_0$. (a) $V_{0}=2\Gamma_0$, (b) $10\Gamma_0$, (c) $20\Gamma_0$, and (d) $50\Gamma_0$.
A strong coupling $E_J=500\Gamma_0$ between the island and the bulk superconductor is considered. Other parameters:
$E_c=30\Gamma_0$, $\varepsilon_M=0$, $k_{B}T=5\Gamma_0$, $\lambda_{1,2}=10\Gamma_0$, and $n_g=0$.}
\label{fig-7}
\end{figure}

\begin{figure}[htbp]
\centering
\includegraphics[width=0.45\textwidth]{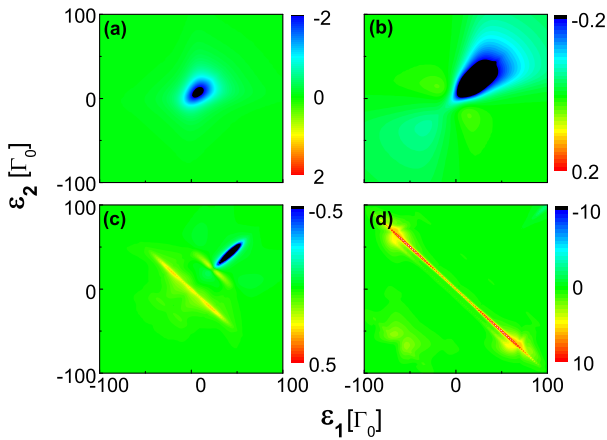}
\caption{The zero-frequency current noise
cross correlation $S_{12}$ as a function of
the dot energy levels $\varepsilon_1$ and $\varepsilon_2$
for different lead chemical potentials $\mu_1=\mu_2=V_0$. (a) $V_{0}=2\Gamma_0$,
(b) $20\Gamma_0$, (c) $50\Gamma_0$, and (d) $100\Gamma_0$.
The coupling between the island and the bulk superconductor is assumed to be comparable to
the charging
energy by taking $E_J=50\Gamma_0$ and $E_c=30\Gamma_0$.
Other parameters:
$\varepsilon_M=0$, $k_{B}T=5\Gamma_0$, $\lambda_{1,2}=10\Gamma_0$, and $n_g=0$.}
\label{fig-8}
\end{figure}

The quantum dot offers an approach to control and modulate the MBSs
and related nonlocal transport by tuning the dot energy levels.
Figures \ref{fig-7} and \ref{fig-8} demonstrate the current noise
cross correlation $S_{12}$ as a function of the dot energy levels
$\varepsilon_1$ and $\varepsilon_2$ for different chemical
potentials $\mu_1=\mu_2=V_0$ in the strong and weak $E_J$ limits,
respectively. It is shown that for different $V_0$, the cross
correlation shows distinct correlation patterns. For a small $V_0$,
the currents are mainly contributed by the thermal activation, and
only a negative cross correlation is induced and centered near
$\varepsilon_i=0$. This feature is different from the correlation
property induced by the Majorana energy splitting, in which a
four-peak cloverlike pattern of noise cross correlation appears when
tuning the dot energy levels \cite{Lu12PRB,Zocher13PRL,Lu14PRB}.
With the increase of $V_0$, the cross correlation $S_{12}$ becomes
positive gradually along the line of $\varepsilon_1=-\varepsilon_2$.
The antisymmetric level configuration tends to enhance the crossed
Andreev reflection processes. As $V_0$ increases further, $S_{12}$
in the region of $\varepsilon_{1,2}< 0$ becomes positive. The sign
reversal of the cross correlation occurs in the region of
$\varepsilon_{1,2}> 0$ when $V_0$ is large enough. In this regime,
the transport channels through all eigen energy levels are open and
the competition between different tunneling paths leads to the
positive $S_{12}$ in all regions. In this case, the cross
correlation $S_{12}$ in the region of antisymmetric level
configuration ($\varepsilon_1=-\varepsilon_2$) is much stronger than
those in other regions. When the coupling strength $E_J$ is
comparable to the charging energy $E_c$, the enhanced current cross
correlations appear in the large $V_0$ limit, as a result of the
dynamical Coulomb blockade effect. When $E_J$ is strong enough, all
the Cooper pair number states $|N_c\rangle$ are well mixed, which
suppresses the Coulomb blockade effect. Therefore, the positive
cross correlation in the strong $E_J$ limit becomes much weaker.

\begin{figure}[htbp]
\centering
\includegraphics[width=0.45\textwidth]{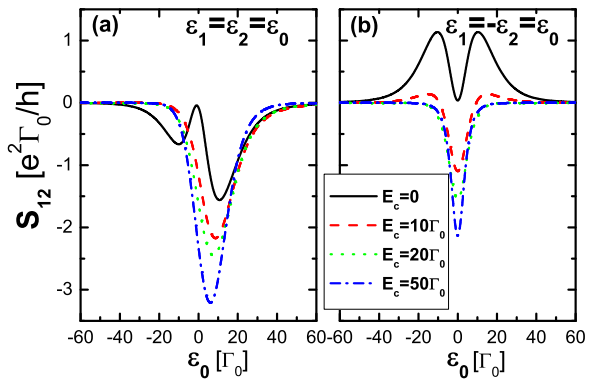}
\caption{The zero-frequency current noise
cross correlation $S_{12}$ at (a) symmetric ($\varepsilon_1=\varepsilon_2$) and (b) antisymmetric
($\varepsilon_1=-\varepsilon_2$) level configurations of two dots
for different intra-island charging energies $E_c=0$, $10\Gamma_0$, $20\Gamma_0$, and $50\Gamma_0$.
A finite Majorana energy splitting $\varepsilon_M=20\Gamma_0$ is considered.
Other parameters: $E_J=100\Gamma_0$, $V_{0}=2\Gamma_0$,
$k_{B}T=5\Gamma_0$, $\lambda_{1,2}=10\Gamma_0$, and $n_g=0$.}
\label{fig-9}
\end{figure}

Above we consider the case of exact Majorana zero mode
($\varepsilon_M=0$) to ensure that the current cross correlation is
purely induced by the charging energy in the superconducting island.
Now we turn to investigate the effect of a finite Majorana energy
splitting. In the noninteracting case and at small voltages,
previous studies \cite{Lu12PRB,Zocher13PRL,Liu13PRB,Lu14PRB} have
shown that a finite Majorana energy splitting induces a four-peak
cloverlike pattern of cross correlation with two positive parts (for
$\varepsilon_1\varepsilon_2<0$) and two negative parts (for
$\varepsilon_1\varepsilon_2>0$) when tuning the two dot energy
levels. Figure 8 shows the current noise cross correlation $S_{12}$
as a function of dot energy levels for the case of symmetric
($\varepsilon_1=\varepsilon_2$) and antisymmetric
($\varepsilon_1=\varepsilon_2$) level configurations of two dots at
small lead voltages. We take $\varepsilon_M=20\Gamma_0$ and
different intra-island charging energies. It is illustrated in Fig.
\ref{fig-9} that a finite energy splitting $\varepsilon_M$ favors to
suppress the negative cross correlation. In the noninteracting case
$E_c=0$, the cross correlation is negative (positive) for the
symmetric (antisymmetric) level configuration, consistent with the
previous studies \cite{Lu12PRB,Zocher13PRL,Liu13PRB,Lu14PRB}. If
$E_c$ is comparable to $\varepsilon_M$, the positive cross
correlation is suppressed almost completely, while the negative
cross correlation is enhanced. It is shown in Fig. \ref{fig-9} that
a small charging energy $E_c$ could smear the four-peak cloverlike
correlation pattern formed in the noninteracting case.

\subsection{Comparison with a non-Majorana setup}
For comparison, we also consider a non-Majorana setup. It consists
of a superconductor island coupled to two quantum dots, and each dot
is connected to a normal metallic electrode. In the case that the
pairing strength in the superconducting island is much stronger than
other system parameters, the quasiparticles in the superconductor
are inaccessible, and one can trace out of the degrees of freedom of
the superconducting island without inducing any dissipative
dynamics. An effective Hamiltonian to describe the dynamics of the
double dots could be obtained by performing a real-time perturbative
expansion \cite{Eldridge10PRB,Governale08PRB}. In the limit of an
infinite intradot Coulomb interaction, the effective model
Hamiltonian of the dots-superconducting island part is given by
\begin{eqnarray}
H_{\mathrm{eff}}&=&\varepsilon_1d_1^\dagger d_1+\varepsilon_2d_2^\dagger d_2
+\Delta_{\text{eff}}(d_{1}^\dagger d_{2}^\dagger e^{-i\varphi}+d_2d_1e^{i\varphi})
\nonumber\\
&&+E_c(\hat{N}_c-n_g)^2-E_J \cos \varphi.
\label{Heff}
\end{eqnarray}
Here $\Delta_{\text{eff}}$ is the effective pairing strength between the
quantum dots, and it is determined by the coupling strength between
the superconducting island and quantum dots. The pairing terms in
$H_{\text{eff}}$ describe the formation of nonlocal superconducting
correlations between the two dots induced by the splitting of Cooper
pairs into the two dots.

\begin{figure}[htbp]
\centering
\includegraphics[width=0.45\textwidth]{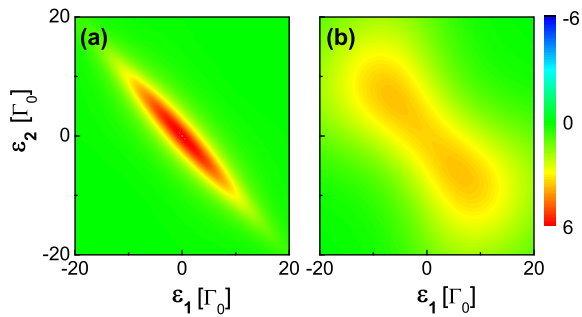}
\caption{The zero-frequency current noise
cross correlation $S_{12}$ as functions of
dot energy level $\varepsilon_1$ and $\varepsilon_2$
for (a) $E_J=20\Gamma_0$ and (b) $E_J=500\Gamma_0$ in the absence of the MBSs.
Other parameters: $E_c=30\Gamma_0$, $V_{0}=2\Gamma_0$,
$k_{B}T=5\Gamma_0$, $\lambda_{1,2}=10\Gamma_0$, $n_g=0$, and
$\Delta_{\text{eff}}=10\Gamma_0$.}
\label{fig-10}
\end{figure}

Here we neglect the processes that electrons tunnel via the
superconductor from one dot to the other, which involves virtual
occupation of quasiparticle states above the gap. It should be noted
that this device can still host MBS on each dot if the electron
tunneling processes between two dots are considered
\cite{Leijnse12PRB}. Meanwhile, an inhomogeneous magnetic field is
considered to make two dots fully spin-polarized, but in different
directions with an angle $\theta$. The amplitude for tunneling
between the dots therefore depends on the angle $\theta$ as $t =
t_0\cos(\theta/2)$, where $t_0$ is the tunneling amplitude for
parallel fields. Similarly, the effective pairing
$\Delta_{\text{eff}}$ also depends on the angle and is given by
$\Delta_{\text{eff}}=\Delta_{0}\sin(\theta/2)$, where $\Delta_{0}$
is the pairing strength between two dots for antiparallel spin
polarizations. In the case of $t=\Delta_{\text{eff}}$ and either
$\varepsilon_1$ or $\varepsilon_2$ being equal to zero, zero energy
solutions of MBSs could be obtained \cite{Leijnse12PRB}. When the
superconducting gap is much larger than other energy scales,
electrons tunneling between two dots via the superconductor will be
strongly suppressed. Therefore, the validation of the non-Majorana
setup we considered is that the processes of electrons tunneling
between two dots via the superconductor are rather weak, i.e.,
$t\ll\Delta_{\text{eff}}$.

In the absence of MBSs, there are no other low-energy quasiparticle
states in the superconducting island and electrons tunnel from the
quantum dots into the island in pairs. In the limit of infinitely
strong Coulomb interaction in the quantum dots, the currents are
dominated by the crossed Andreev reflection. Only a positive current
cross correlation could be generated and the sign of the cross
correlation is independent of the dot energy levels and the charging
energy. In Fig. \ref{fig-10} the current noise cross correlation is
plotted as a function of dot energy levels in the absence of MBSs.
Only a positive cross correlation exists for the non-Majorana
device. This is quite different from the results in the presence of
MBSs. In Figs. \ref{fig-6}-\ref{fig-8}, only a negative cross
correlation is generated at low lead chemical potentials. Therefore,
we can check the existence of MBSs from the sign of the current
noise cross correlation at the small lead chemical potentials.

We have assumed that the on-site Coulomb interaction in the two dots
is so strong that only one electron is allowed to stay in the dot at
one time. In this case, the direct Andreev reflection processes are
suppressed. Finally we briefly discuss the effect of the direct
Andreev reflection processes on the current noise cross correlation.
If the two dots are connected to a bulk superconductor, the crossed
Andreev reflection induces positive cross correlation, while the
direct Andreev reflection processes do not contribute to the cross
correlation. However, if the two dots are coupled to a
superconducting island with a finite charging energy, the direct
Andreev reflection processes through one dot will compete with the
tunneling in another dot to lower the energy in the island. In this
case, it is expected that a negative cross correlation is induced.

\section{Conclusion}

In conclusion, we investigate the currents and their noise cross
correlations modulated by a global charging energy in a hybrid
device of MBSs and quantum dots. The MBSs are created in a
semiconductor nanowire in proximity to a mesoscopic superconducting
island. Each end of the nanowire is weakly connected to a quantum
dot and a normal metal lead. This configuration is motivated by the
possibility to switch the sign of current noise cross correlation by
varying the dot energy levels. We have studied two cases: the
superconducting island is floating, and the island is connected to
another bulk superconductor. The floating island case shows a
negative differential conductance and a giant super-Poissonian shot
noise, resulting from the interplay of the nonlocality of the MBSs
and dynamical Coulomb blockade.

When the superconducting island is coupled to the bulk
superconductor, we show that the current noise cross correlation is
always negative as a function of dot energy levels at low lead
chemical potentials. In contrast, the current cross correlation is
always positive in a non-Majorana setup. For current cross
correlations induced by the energy splitting of MBSs, a four-peak
cloverlike pattern of the noise cross correlation appears when
tuning the dot energy levels. However, the pattern becomes more
complex in the presence of a charging energy on the superconducting
island. With the increase of lead chemical potentials, the cross
correlation becomes positive gradually for the antisymmetric
configuration of the dot energy levels. When the coupling strength
between the island and the bulk superconductor is comparable to the
charging energy, the noise cross correlation is strongly enhanced in
the antisymmetric level configuration. The sign of the current cross
correlation in the low lead chemical potentials can serve as a
signature for the existence of the MBSs.

This work was supported by the Research Grant Council, University
Grants Committee, Hong Kong under Grant No. HKU703713P and the Natural
Science Foundation of China under Grant No. 61474018 and No.
11574127.

\end{document}